\documentclass[manuscript]{aastex}

\usepackage{amsmath}

\slugcomment{Not to appear in Nonlearned J., 45.}

\shorttitle{Seris}
\shortauthors{De la Luz et al.}

\begin{document}

\title{Synthetic Spectra of Radio, Millimeter, Sub-millimeter and Infrared Regimes with NLTE approximation}

\author{Victor De la Luz\altaffilmark{1,2}, 
Alejandro Lara\altaffilmark{1} and Jean-Pierre Raulin\altaffilmark{3}}
\altaffiltext{1}{Instituto de Geof\'isica, Universidad Nacional Aut\'onoma de
  M\'exico, M\'exico, 04510.}
\altaffiltext{2}{Instituto Nacional de Astrofisica, Optica y Electronica,
  Tonantzintla, Puebla, Mexico, Apdo. Postal 51 y 216, 72000.}
\altaffiltext{3}{CRAAM, Universidade Presbiteriana Mackenze, S\~ao Paulo, SP, Brasil, 01302-907.}


\begin{abstract}
We use a numerical code  called PAKALMPI  to
compute  synthetic spectra of the solar emission  
in quiet conditions
at millimeter, 
sub-millimeter and infrared wavelengths. 
PAKALMPI solves the radiative transfer equation, with Non Local
Thermodynamic Equilibrium (NLTE),  in a three dimensional 
geometry using a multiprocessor environment. 
%
%
The code is able to use three opacity functions: classical bremsstrahlung, 
$H^-$ and inverse bremsstrahlung. 
In this work we have computed and compared two
synthetic spectra, one in the common way: 
using bremsstrahlung  opacity function and considering a 
fully ionized atmosphere; and a new one considering 
bremsstrahlung,  inverse bremsstrahlung and $H^-$ opacity 
functions in NLTE. 
We analyzed in detail the local behavior of the low atmospheric emission 
at 17, 212, and 405 GHz (frequencies used by the Nobeyama
Radio Heliograph and the Solar Submillimeter Telescope).
We found that the $H^-$ is
the major emission mechanism at low altitudes (below 
500 km) and
that at higher altitudes the classical bremsstrahlung becomes the major 
mechanism of emission. However the brightness temperature remains unalterable.
Finally, we found that the inverse bremsstrahlung process is not important 
for
the radio emission at these heights.

\end{abstract}


\section*{Introduction}

The radio emission from astronomical sources may be a powerful tool to study
the physical conditions of the source and the medium between it and the observer.
In particular 
at centimetric, millimetric and sub-millimetric wavelengths 
the radio emission from the
solar atmosphere 
carries out valuable information from the 
physical conditions of the
 chromosphere and 
the transition region.

The solar emission at high radio frequencies can
be explained 
by
two major solar regimes:
the flaring case, where 
non-thermal emission from accelerated electrons and thermal emission from hot
and cold plasmas are the main processes;
and the quiet Sun regime, where the emission
is attributed to thermal bremsstrahlung \citep{1970SoPh...13..348K}, 
for which 
the magnetic field is not directly important. Notice that 
the magnetic field 
structures the matter in the
solar atmosphere, therefore it will influence the plasma density structure, 
as instance, 
in the case of  chromospheric spicules.

In the early 1930's, the first attempts were made to explain the 
solar chromosphere \citep{1935HarCi.410....1C}.
But it was only after the eclipse observations of 1952, that 
the existence of a two-component model in temperature was established 
\citep[hot and  cold temperature   model,][]{1949MNRAS.109..298G}. 

The classical treatment of 
the solar chromosphere in quiet regime is based on the assumption
that the atmosphere is static and in hydrostatic equilibrium
\citep{1973ApJ...184..605V}.
Also, the radio and UV spectra must be in agreement, therefore 
in several works have been 
studied the relationship between the UV line emission and the continuum at
 millimeter and sub-millimeter wavelengths 
\citep{1978SvA....22..345K,
  1981SoPh...69..273A,1983SoPh...85..237C,
  2003ApJ...589.1054L,2004A&A...419..747L,
  2004A&A...422..331C,2008GeofI..47..197D}.

By assuming an emission mechanism, 
the observed radio-spectrum  
can be correlated with the atmospheric
height of the source
\citep{1976ApJS...30....1V,1989SoPh..124...15A,2008ApJS..175..229A,2010ApJS..188..437D}. 
Generally, the height associated to a specific emitting frequency is the height
of the layer 
where the atmosphere becomes optically thick. 
However,
the transition between optically thin ($\tau_{\nu}\ll 1$) 
and optically thick regimes ($\tau_{\nu} \gg 1$) is quite ambiguous.

Theoretically, the solar emission can be modeled by solving the radiative
transfer equation taking into account the temperature and density 
of the solar atmosphere, 
and using the required opacity functions 
according with the relevant photon-particle
interactions.
%
In particular, from  
millimeter to infrared wavelengths
range, 
the free-free
interaction (bremsstrahlung) is the main  mechanism used to
reproduce 
solar spectra \citep{1985ARA&A..23..169D,2005AdSpR..35..739R},
and, also, a fully ionized atmosphere is commonly assumed.
With these 
assumptions both the time and the complexity of the 
computations are greatly reduced, specially for the quiet Sun emission case.

It is clear that more  photon-particle interactions take place in the solar 
atmosphere, in particular, 
the $H^-$ continuum 
provides the dominant radiation  at photospheric level 
\citep{1961psc..book.....A}, and therefore $H^-$ must
be considered as an important source of sub-millimeter and infrared emission 
at low chromospheric level. 
We expect  that the 
interactions between $H^-$ and free electrons
\citep[inverse bremmstrahlung, ][]{1980ZhTFi..50R1847G} is  important at this wavelenght range also
.

Therefore, in order to 
reproduce a more realistic solar 
spectra, 
it is necessary to compute
explicitly 
the density of the different ion species
and  feed with this information the relevant opacity functions, 
using Non Local Thermodynamical Equilibrium (NLTE).
%
%
In this way,
we have constructed a synthetic 
spectrum
of the solar
quiescent
radio emission at short wavelengths.

%
As a first step, we  compute
the ion and 
electron density at any layer of the
atmosphere using as inputs, atomic models and pre-calculated 
density and temperature atmospheric
profiles
for the quiet Sun regime.
Then,
we
solve the radiative 
transfer equation for the required ray paths.  
We use PAKALMPI, an updated version of the PAKAL code \citep{2010ApJS..188..437D} to
compute a synthetic spectrum
of the solar emission from 
millimeter to infrared wavelengths.
PAKALMPI solves the radiative transfer equation, in 
NLTE conditions, in a three dimensional 
geometry using a multiprocessor environment. %

The computed synthetic spectra can be used to study the detailed
behavior of the emission and  absorption throughout the solar
atmosphere. In particular, in this work we are interested in the amount 
of local emission at different atmospheric heights for the millimeter -
infrared  
wavelength range. We focus at 
17, 212, and 405 GHz, because these are the observing frequencies of
the Nobeyama Radio Heliograph \citep[NoRH,][]{1995JApAS..16..437N} and the Solar Submillimeter Telescope 
\citep[SST,][]{2008SPIE.7012E..19K}. For this, 
we use NLTE computations for Hydrogen and $H^-$ ion, taking into
account three opacity functions (bremsstrahlung, $H^-$ and inverse
bremsstrahlung) for twenty ion species in different ionization stages.

The goal of this work is to compare
the common
synthetic spectra
obtained by
using the bremsstrahlung  opacity function only with 
fully ionized atmosphere, and a more realistic spectra calculated by using
NLTE and three opacity functions, computed with high spatial resolution
(of around 1 km).

The three opacity functions, the computation of ionization stages of
 twenty ion species considering $H$, $H^-$, and $n_e$ in NLTE are
included in our new model called ``Celestun''. 


\section{Opacity Functions}
%
In order to include the contribution of  
different ion species, 
we use the following (complete) opacity function which resumes the 
slightly different expressions for the bremsstrahlung opacity reported
in the literature
%
\citep{1979ApJS...40....1K, 1985ARA&A..23..169D,1986rpa..book.....R, 1996ASSL..204.....Z}:
\begin{equation}
\kappa_{ff}^{n}(T,\nu) = \sum_\xi\frac{2^{5/2}\sqrt{\pi}}{3\sqrt{3}}\frac{e^6}{c(mk)^{3/2}}\frac{Z_\xi^2}{T^{3/2}\nu^2}n_en_\xi\bar{g}_{ff}(Z_\xi,T,\nu),
\end{equation}
where: $e$ is the electron charge; $c$ is the velocity of light; 
$m$ the electron mass; $k$ the Boltzmann constant; $Z_\xi$ is the 
charge of the ion species $\xi$; $T$ is the electron
temperature; $\nu$ the frequency; 
$n_e$ the electron density; $n_\xi$ is the density of
the ion species $\xi$; and
$\bar{g}_{ff}(Z_\xi,T,\nu)$ is the average free-free Gaunt factor
which
takes into account the initial and final quantum energy states
of the total momentum. 

The rigorous quantum mechanical expression for the Gaunt factor  
has been derived by 
\cite{1935AnP...415..589S} but the solution 
is difficult to compute and therefore,  many approximations have been developed
in order to make the computation easier (see \cite{1962RvMP...34..507B} 
for 
%
a review of
average Gaunt factor in the bremsstrahlung opacity function).
%

Following \cite{1986rpa..book.....R} we 
compute the average free-free Gaunt factor for  chromospheric
conditions as:
\begin{equation} \label{laopacidad}
\bar{g}_{ff}(T,Z,\nu) =  
\begin{cases} 
\frac{\sqrt{3}}{\pi} \ln \left[\frac{4}{\xi^{5/2}y}x^{1/2}\right] & x < 1, y < g(x) \\
1 , & x \le 1 , y > g(x),  y \le h(x)\\
\left[\frac{12}{xy}\right]^{1/2} & x \le 1,  y > h(x)\\
\frac{\sqrt{3}}{\pi}\ln\left[\frac{4}{y\xi}\right] & x >1, y < 1\\
\left[\frac{3}{\pi y}\right]^{1/2} & x > 1, y \ge 1
\end{cases}
\end{equation}
where
\begin{equation}
x =  \frac{kT}{Z^2Ry},
\end{equation}
and $Ry=13.6\ \mbox{eV}$ is the Rydberg constant and
\begin{equation}
y = \frac{h\nu}{kT},
\end{equation}
\begin{equation}
g(x)= \frac{9999}{10^4}\left(\frac{99}{10^2}x+ \frac{99}{10^3}\right),
\end{equation}
\begin{equation}
h(x)= 1001-1000x.
\end{equation}
Under chromospheric conditions it is necessary to include 
the contribution of the $H^-$ ion in the opacity expression.
This  contribution is effective through
three mechanisms 
\citep{1988A&A...193..189J,1980ZhTFi..50R1847G}:

\begin{itemize}
\item The Wildt's photo-detachment mechanism \citep{1939ApJ....90..611W}
$$
h\nu + H^- \rightarrow H +e^-
$$
\item The neutral interaction \citep{1996ASSL..204.....Z}
$$
h\nu + e^- + H \rightarrow H + e^-
$$
\item The bremsstrahlung for $H^-$
or inverse bremsstrahlung \citep{1980ZhTFi..50R1847G}
$$
h\nu + e^- + H^- \rightarrow H^- + e^-.
$$
\end{itemize}

In the millimeter - sub-millimeter wavelength range, 
the photo-detachment
mechanism is not 
relevant
\citep{1994ApJ...437..879A}, 
whereas the neutral interaction
is very important. 
The corresponding opacity function is
$$
\kappa_{ni}^{H} = k_{\nu}^{tot}P_en_H, 
$$
where $k_{\nu}^{tot}$ is the total absorption coefficient published
in \cite{1988A&A...193..189J}, $P_e$ is the electronic pressure and
$n_H$ is the Hydrogen density.
Finally, in the wavelength range of interest,
the inverse bremsstrahlung  opacity
is the same than the positive ion case 
\citep[we use bremsstrahlung with $n_i = n_{H^-}$]{1980ZhTFi..50R1847G}.

Therefore, the total opacity used to compute the synthetic spectrum is
$$\kappa_{ff} = \kappa_{ff}^{n} + \kappa_{ni}^{H} + \kappa_{ff}^{H^-}.$$
%
\section{Density Profiles}

As input parameters 
 PAKALMPI needs the Hydrogen 
density, the temperature, and the departure coefficients (see Sec. \ref{nlte}).
%
With this information, PAKALMPI computes the ionic and electronic densities 
following 
the derivation
of \cite{1973ApJ...184..605V}, but using a higher number of energy levels
for each ion,
i.e 
the Saha equation is solved not only for the base state. 

We define
the ionization contribution function ($Z$ function) by:
\begin{equation}\label{Zorro}
Z=\sum_{\xi = He}^{\Xi}\sum_{N_{\xi}=1}^{F} N_\xi n_{\xi,N_\xi},
\end{equation}
where 
$$
\xi = \{H,He, Li, Na, K, \dots, \Xi\}.
$$
are the atoms considered in the model; $N_{\xi}$ is the ionization level;
$n_{\xi,N_\xi}$ is the density of the atom $\xi$ in the ionization
level $N_\xi$; and F is the highest energy level for each ion. 
In order to calculate the Z function 
we need an atomic model 
and the relative density or abundance of each atom. Although the relative 
density may depend of the 
atmospheric height,
in a first approximation, 
we assume
a constant relative density for all the 
atmospheric layers.
\subsection{Electron Density}
In \cite{1973ApJ...184..605V} we found the formulas for the electron density
which takes into account NLTE computations and uses the
fundamental energy
state for the
ions. However, using our definition of ionization contribution function
(Eq. \ref{Zorro}), 
we have:
\begin{equation}\label{elecde}
n_e= \frac{-(1-Zd)+\sqrt{(1-Zd)^2+4d(n_HZ)}}{2d},
\end{equation}
where $n_H$ is the Hydrogen density, $Z$ is the ionization contribution 
function, $T$ the temperature and $d$ the  
NLTE contribution function.
%
%
%

\subsection{Ion Density}
To compute $n_{\xi,N_\xi}$, we solve the  matrix for the 
classical Saha equation:
\begin{equation}\label{saha}
\log\frac{n_{\xi,N_{\xi+1}}}{n_{\xi,N_{\xi}}} = -0.1761-\log(P_e)+\log\frac{u_{N_{\xi+1}}}{u_{N_{\xi}}}+2.5\log T -\chi_{\xi,N_{\xi},k}\frac{5040}{T},
\end{equation}
where
\begin{equation}
u_{N_{\xi}} = \sum_n g_{\xi,N_{\xi}}\exp{\left(-\frac{\chi_{\xi,N_{\xi},k}}{kT}\right)}.
\end{equation} 

PAKALMPI can use a single model of atoms, but is able to manage
different atomic densities in each layer.
Even more, it is possible to change the metalicity in each layer of the 
atmosphere.
In this work, we use
the atomic model  
 published by
\cite{1973ApJ...184..605V}.

To calculate the density of $H^-$ we use the following expression \citep{1973ApJ...184..605V}:
$$n_{H^-} = 1.0345\times 10^{-16}b_{H^-}n_en_{HI}T^{-3/2}e^{8762/T}.$$


\subsection{NLTE Contribution Function}\label{nlte}
The NLTE 
contribution function is:
\begin{equation}\label{defd}
d=b_1\psi(T)\left(1+\sum_{l=2}^N\frac{n_l}{n_1}\right),
\end{equation}
where the term $ \sum_{l=2}^N\frac{n_l}{n_1} << 1$ 
\citep{1973ApJ...184..605V}, so that,
\begin{equation}\label{defd2}
d=b_1\psi(T),
\end{equation}
where
\begin{equation}
\psi(T) = \left(\frac{h^2}{2\pi mkT}\right)^{3/2}\exp(\chi_H/kT),
\end{equation}
and the departure coefficient for the Hydrogen in the fundamental state 
($b_1$)
 defined in \cite{1937ApJ....85..330M} and modified by \cite{1973ApJ...184..605V}
is:
$$
b_1 = \frac{n_1/n_1^*}{n_k/n_k^*},
$$
where the ratio $n_1^*/n_k^*$ is given by the Saha equation in thermodynamical
equilibrium.

As $b_1$ 
is difficult to compute for each specific atmospheric layer
and due to the fact that 
there are only
few published atmospheric models reporting this parameter, 
we have developed an algorithm 
to obtain 
an approximation of
the departure coefficient
based on 
any 
published $b_1$ coefficients. 

We found that different 
theoretical models of the chromosphere 
are very close to each other
when plotted in 
the dynamic space (density versus temperature,  
the analysis in
the dynamic
space is important because in this space, the problem of the ambiguity between 
height and temperature is avoided, 
this is, when 
the temperature profile has two or more associated heights for a single 
temperature point).
Therefore, it should be easy to get new approximated $b_1$ parameters
from published departure coefficients.
If we have two models:
a published model (model A) of the atmosphere with: 
temperature $T_A(h)$, density $H_A(h)$, and departure coefficient 
parameter $b_1(h)$, where $h$ is the height over the photosphere; and
if we have a new model (model B) with  temperature $T_B(h)$ 
and density $H_B(h)$; we can approximate $b_1$ from model A 
as follows:
\begin{enumerate}
\item At a given height over the photosphere ($h_B$) we take the temperature and
density from the model B 
$$\vec{p_B} = (T_B(h_B),H_B(h_B)).$$
\item With $\vec{p_B}$ we found the closer point 
$\vec{p_A} = (T_A(h_A),H_A(h_A))$ to the
model A (in the dynamic space). This point has an associated height ($h_A$) 
over the photosphere in the model A.
\item We take the points that enclose $h_A$ 
$$[h_0,h_A , h_1]$$
in the same  model A.
\item With $h_0$ and $h_1$ we can evaluate $b_1$ in the model A in order to 
interpolate $h_A$:
\begin{equation}
b_1(h_B) \approx b_1(h_A) = \frac{b_1(h_1)-b_1(h_0)}{h_1-h_0} (h_A-h_0) +b_1(h_0).
\end{equation}
\end{enumerate}





 



By construction, $\vec{p_A}$ and $\vec{p_B}$ have similar physical conditions.
Therefore, 
the computed departure parameter 
 $b_1(h_B)$ 
can be used in the model B with high degree of confidence.
 This process is applied to all  points  in model B to
obtain the required approximation for the departure parameters.


\section{Radiative Transfer Equation}
The local emission, or the emission in a given atmospheric layer, 
can be represented in the following way
\begin{equation}
I_{local} = \epsilon_{abs} + \epsilon_{emi}
\end{equation}
where 
\begin{equation} \label{eq:eabs}
\epsilon_{abs} = I_0\exp(-\tau_{local})
\end{equation}
is the local absorption  and 
\begin{equation}\label{eq:eemi}
\epsilon_{emi} = S_{local}(1-\exp(-\tau_{local}))
\end{equation}
is the local emissivity. 

In this case $I_0$ is the incoming emission
from the background (lower adjacent atmospheric layer), $\tau_{local}$ is the local optical depth and $S_{local}$ is
the local source function.

We  use an updated version of PAKAL code to solve the radiative transfer equation 
\citep{2010ApJS..188..437D} 
along a ray path using an
iterative method to compute $I_0$ and $I_{local}$.

The $\epsilon_{abs}$ and $\epsilon_{emi}$ are useful to analyze the 
behavior of the local emissivity of the atmospheric layers. 
While $\epsilon_{abs}$ is related to
the level of opacity of a given layer, 
$\epsilon_{emi}$
indicates
the capacity of this layer to produce radiation.

With this definition of the radiative transfer equation, we can
study in detail the local emission and absorption processes.

\section{Algorithm} 
The code PAKALMPI performs the following steps to compute the emission $I_\nu$:
\begin{itemize}
\item
Reads the temperature, Hydrogen density, and the atomic models for 
a given layer 
from the atmospheric pre-calculated model (VALC,
C07, etc),
\item
Solves the matrix of the Saha equation (Eq. \ref{saha}). The solution of the
matrix is a set of density of ions for each atomic species considered 
in this layer.

\item
Finds the ``Z'' contribution of electrons in LTE (Eq. \ref{Zorro}). 
\item
Computes the approximation to the $b_1$ parameter using the temperature 
and density of the specific layer.
\item
Using  $b_1$ and $Z$, computes the initial values of $n_e$ (Eq. \ref{elecde})
and $n_i$. 
\item
Computes the $n_{H^-}$ and recalculates 
the $n_e$ and $n_{HI}$  (this step is necessary since the computation of 
$n_{H^-}$ 
needs the electron and neutral Hydrogen densities).
\end{itemize}
These  steps are repeated recursively 
until $n_e$ converges. The iterative process 
 is necessary because we are searching for the equilibrium of the physical system.

Once $n_e$ has converged, the results for all ions are saved and the code gets the 
temperature, Hydrogen density, and abundances for the next layer.
In this way, the code produces 
tables 
with the 
height profiles
of the electron and ion densities.

When all the atmospheric layers 
have been calculated,  
PAKALMPI computes the ray trajectory (
from the Sun to the observer) 
and the intersection of this ray path with the radial models 
(ions, temperature and electronic density) 
from the computed tables. Finally, PAKALMPI  
 solves the radiative transfer equation using all the considered opacity 
functions.


The process is repeated (for each ray trajectory) to create 2D images at several 
frequencies. In this way, PAKALMPI improves the construction of 2D images 
considering 
detailed 3D emitting structures.


\section{Chromospheric Spectra}




As inputs to our code, we use two complementary atmospheric models, 
the  chromospheric 
model C07  \citep{2008ApJS..175..229A} which 
includes temperature and Hydrogen density but does not include the departure coefficients, and the VALC model
\citep{1981ApJS...45..635V} in order to get these departure coefficients.

\begin{figure}
\begin{center}
\includegraphics[width=1.0\textwidth]{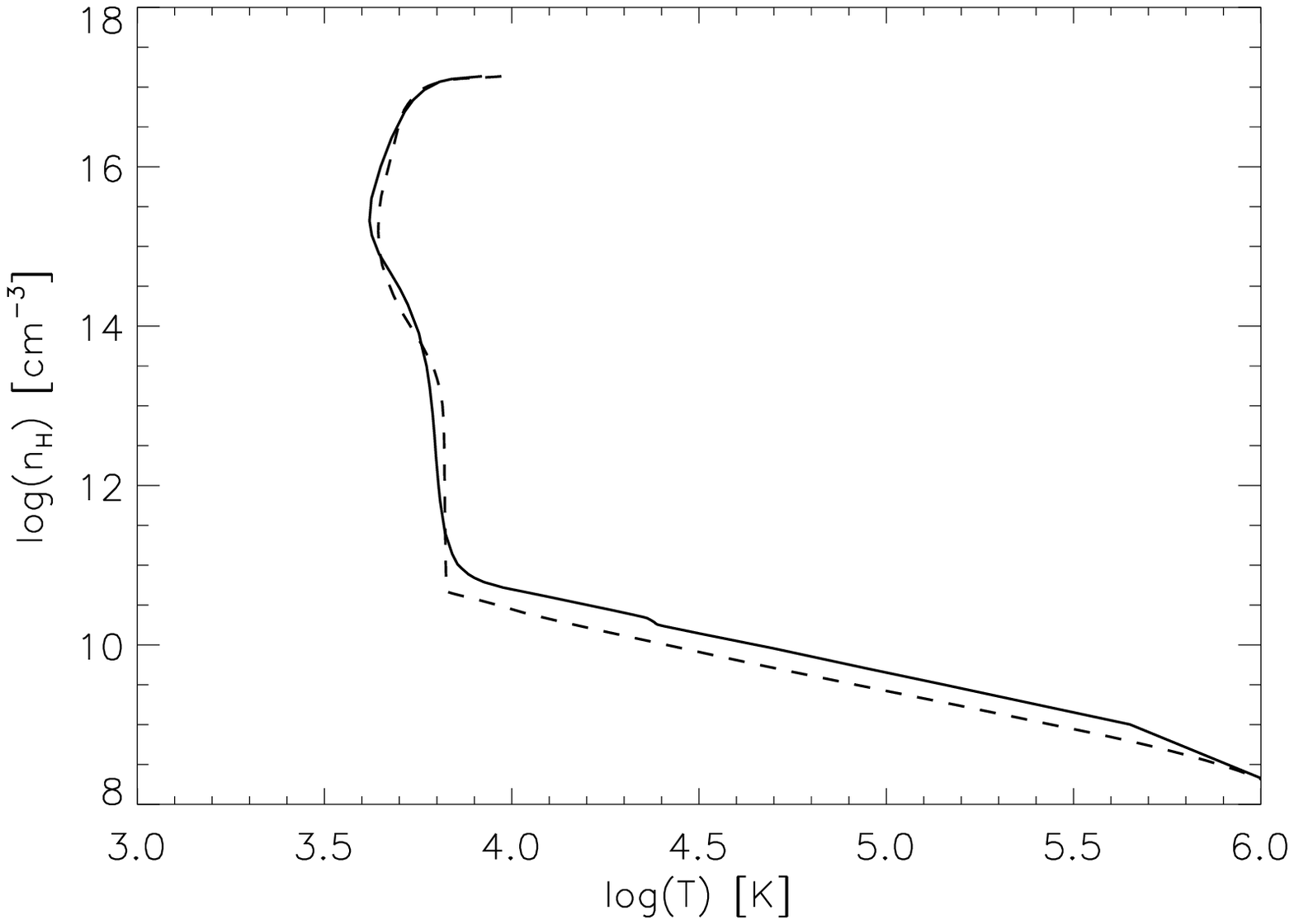}
\caption{Temperature vs Hydrogen density for
VALC model (solid line) and C07 Model (dashed line)
in the dynamic space.
}\label{figure1.ps}
\end{center}
\end{figure}

\begin{figure}
\begin{center}
\includegraphics[width=1.0\textwidth]{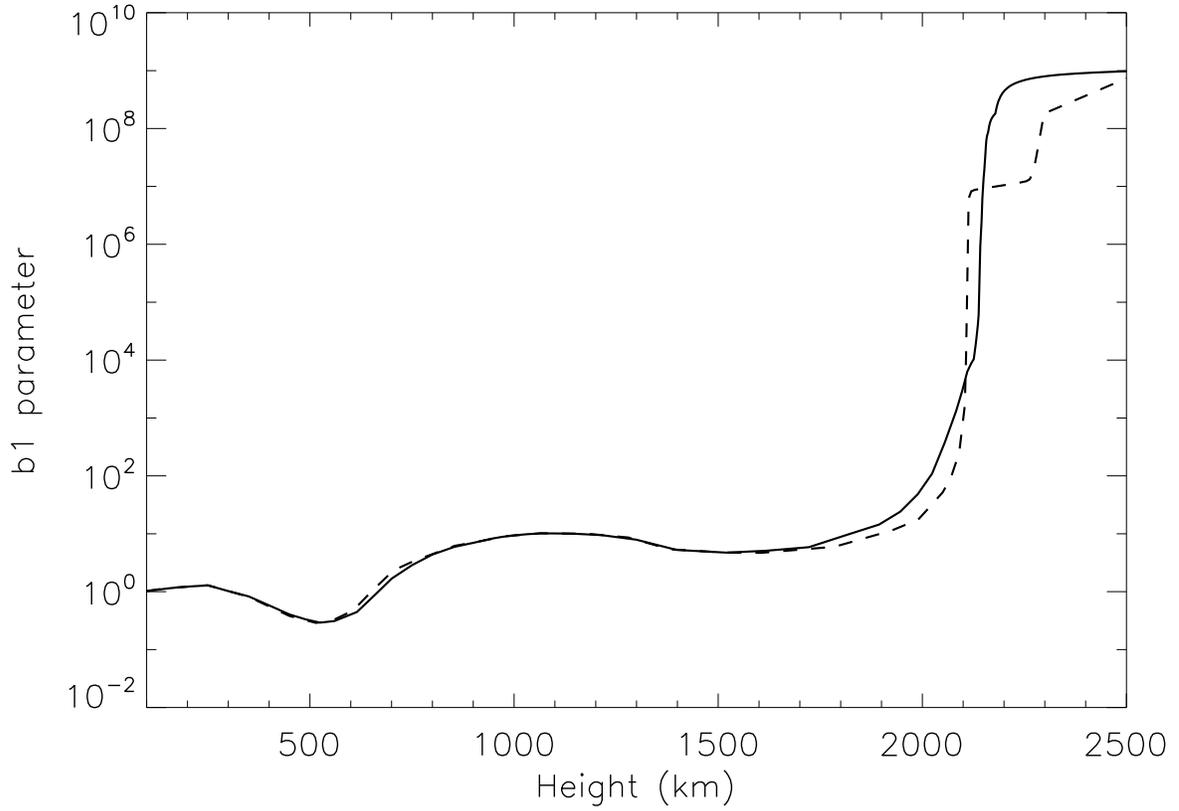}
\caption{Departure coefficients for VALC (dashed line) and C07 
(continue line) approximated model
taking into account the $b_1$ parameter published in
the VALC  model. We 
note
that 
for
both 
models
$b_1$ parameter follow the radial
temperature. The main differences are seen above 1500 km over the photosphere.
}\label{figure2.ps}
\end{center}
\end{figure}

The (atmospheric) plasma behavior can be easily visualized by plotting the plasma
density versus the temperature, this is the dynamic space. Therefore,    
in  Figure \ref{figure1.ps} we have plotted the neutral Hydrogen 
density as a function of the temperature for  VALC 
and C07 models. 
We note
that the differences between both models are 
minimal.
This gave us confidence to 
 use both: the C07 atmospheric model in order to get Hydrogen 
density and temperature; and 
%
the departure coefficients for the 
fundamental
state of  $H$ ($b_1$) 
and $H^-$  ($b_{H^-}$) 
which 
are
interpolated from VALC model using our approximation method (Sec. \ref{nlte}). 
The result
is seen
in  Figure \ref{figure2.ps},
where we have plotted the original departure coefficients (VALC) 
with a dashed 
line
and our interpolation (for C07) 
with a continuous line.
%
%
%
%
%
%
%
The main differences correspond to
the region of the 
temperature minimum
and  the transition region, 
and
 are caused by temperature differences between
 both models. 



\begin{figure}
\begin{center}
\includegraphics[width=0.9\textwidth]{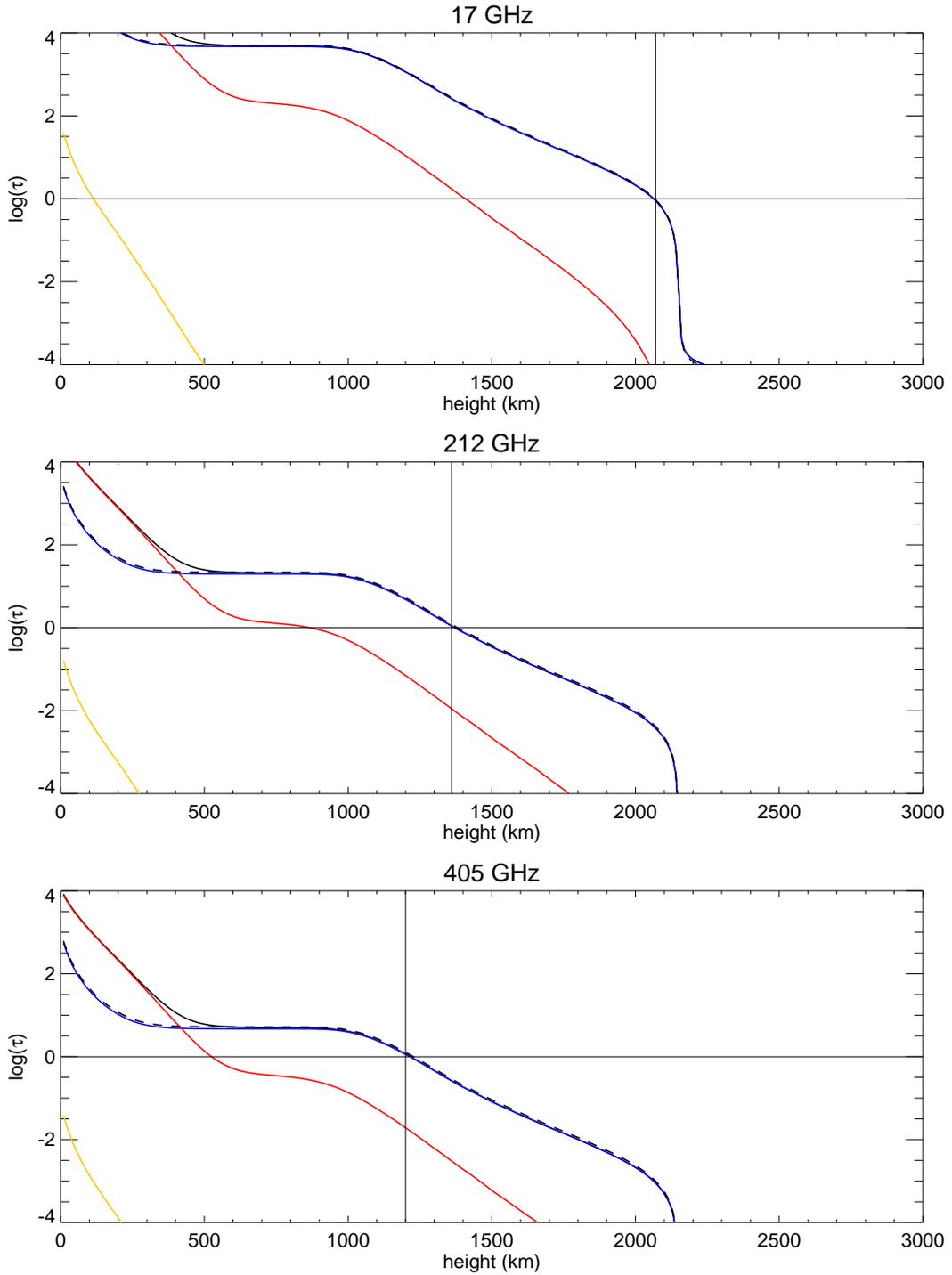}
\caption{Optical Depth at 17, 212, and 405 GHz using the C07 input atmospheric
  model. In blue we 
show
the contribution of the bremsstrahlung opacity, red $H^-$ and yellow the
Inverse bremsstrahlung. The continuous line 
shows the sum of
all the contributions and the dashed line 
represents
only the bremsstrahlung opacity function.}\label{figure5.ps}
\end{center}
\end{figure}

%
Our code output gives us detailed information of the radiation parameters. 
As instance, 
in  Figure \ref{figure5.ps}  we show the computed 
optical depth at 17 (top), 
212 (middle) and 405 (bottom) GHz as a function of radial height.
%
We made two simulations, the first one using only bremsstrahlung
opacity (dashed black lines); 
and the second one using
our  model Celestun, 
in this case,  the yellow lines 
 represent
the inverse bremsstrahlung opacity; 
the $H^-$ opacity is 
shown
by  red lines; 
the blue lines 
correspond
to the classical bremsstrahlung opacity; and  the black
continuous lines represent the full opacity function from Celestun. 
%
%
%
%
The introduction of 
  $H^-$  and  inverse bremsstrahlung opacities  does not 
change the height where  
the atmosphere becomes optically thick ($\tau \sim 1$). 
%
However, it is
clear that 
below
500 km, the $H^-$ process is the
major source of opacity in the 
solar atmosphere.


The detailed local behavior of the emission at 17, 212 and 405 GHz
is
seen  
in  Figure \ref{figure6.ps}, where 
we have plotted  the brightness temperature (black lines)  and the
input atmospheric radial temperature (yellow lines) as a function of the 
atmospheric height. The continuous and dashed lines represent, respectively, 
the cases where all, and
only bremsstrahlung opacities were considered.
The  contribution, to the local brightness,  of the  absorption 
$\epsilon_{abs}$ (Eq. \ref{eq:eabs}) is plotted in blue lines, whereas the 
emission  $\epsilon_{emi}$ 
(Eq. \ref{eq:eemi}) is plotted  in red lines.
%

We note that there are two separated regions where clearly 
the emission has a significant contribution: 
the first one 
is  
close
to the photosphere, and
it is interesting to note that the height of this region 
increases
when 
the $H^-$ opacity function is considered;  
The 
importance 
of the second enhanced emission region decreases with the
frequency, and the $H^-$ opacity does not produce any differences in this region.  
%
The second enhanced region and the $H^-$ opacity do not affect the final 
brightness 
temperature, this is why these structures 
were
unnoticed until now.
The contribution of the inverse bremsstrahlung is negligible.

\section{Synthetic Spectra} 

It is interesting to compare the output of our code using 
different input
atmospheric models. As instance, 
%
in  Figure \ref{figure7.ps}  we show 
observations performed at different phases of the solar cycle and collected by
\cite{2004A&A...419..747L}: diamonds corresponds to close-to-minimum, 
the square symbols to intermediate,
and triangles to close-to-maximum phases of the solar
cycle. In the same figure, we have plotted
the synthetic solar spectra 
computed using all (continuous lines) and only bremsstrahlung (dashed lines) 
opacities  and 
the following input models: VALC in yellow lines; CAIUS05   
\citep{2005A&A...433..365S} in blue lines; 
C7 in black lines; and again CAIUS05 but using their
electronic density and  physical 
assumptions
(bremsstrahlung opacity and fully ionized atmosphere) in red lines.

There are interesting differences in the input atmospheric models, e.g., 
the altitude
and the length 
of the chromospheric  
temperature minimum, as well as 
the altitude and the morphology of the chromospheric upper limit  
(i. e. the beginning of the transition zone and corona). 
The VALC model has the temperature minimum  closer to the
photosphere and is the warmest. This model has a thin
chromosphere and defines a ``plateau zone'', where the chromosphere
temperature reach 20,000 K. This plateau 
is responsible of the 
high brightness temperature at low frequencies  \citep[see Figure 
\ref{figure7.ps}  and][]{2008GeofI..47..197D}.
The CAIUS05 model has similar altitude for the temperature minimum
but is hotter and is shorter compared to the
VALC model. The chromospheric temperature grows exponentially up to  
coronal values. The CAIUS05 model has the largest chromosphere. 
%
The C7 model has the lowest temperature minimum, and then
a constant temperature until the rise due to the transition region and
the corona. This model eliminates the plateau zone
using ambipolar diffusion.
%
%
%
%

In order to facilitate the analysis, 
we consider separately three regions of the spectrum: from 2 to 60 GHz 
(low frequency); 
from 60 to 400 GHz (intermediate frequency); and from 400 GHz to 3 THz 
(high frequency). 

At low frequency range, the temperature differences of the atmospheric
models produce significant differences in the spectra (regardless of the 
opacity function), this emitting region 
is located
at  the upper chromosphere $\sim 1500$ km over the photosphere. 
For intermediate frequencies, the simulations  agree 
rather well with the
observations, however, the observations have large dispersion. 
Finally, at
high frequency range we note that  there is an apparent discontinuity at 400 GHz. 
In this region the model that  better fits the observations is 
CAIUS05. On the other hand, both C07 and VALC models predict higher 
brightness temperatures 
than the observed.

These apparent minimal differences between input atmospheric models, 
produce significant differences in the resultant
synthetic spectra computed from our radiative model.


\section{Discussion and Conclusions}
Using different atmospheric models and opacity functions,
we have computed the optical depth, $\tau$, at low  atmospheric altitudes.
Our results 
show that if we use either, only bremsstrahlung 
or Celestun (i.e. $H$, $H^-$ and inverse Bremsstrahlung) opacities,    
the height over the photosphere where the 
overlying
atmosphere 
becomes optically thick ($\tau \approx
1$) remains unalterable. 
However, if we take into account the local emissivity (Eq. \ref{eq:eemi})
we show that the height where the main emission is  generated, changes
meaningfully. 

For example, at 17 GHz the height over the photosphere where
the atmosphere starts 
to become
optically thin is around $1500$ km, but the
height where $\tau=1$ is around $2050$ km. There are local emissivity 
structures below $1500$ km, although they
remain invisible to the observer because the atmosphere is optically
thick at these heights. 
When we considered bremsstrahlung opacity only, the local emissivity at 17 GHz
changes in a region between $500$ and $850$ km over the photosphere 
(the atmosphere becomes optically thin). 
This structure is not 
present when the Celestun model is considered.

\begin{figure}
\begin{center}
\includegraphics[width=0.9\textwidth]{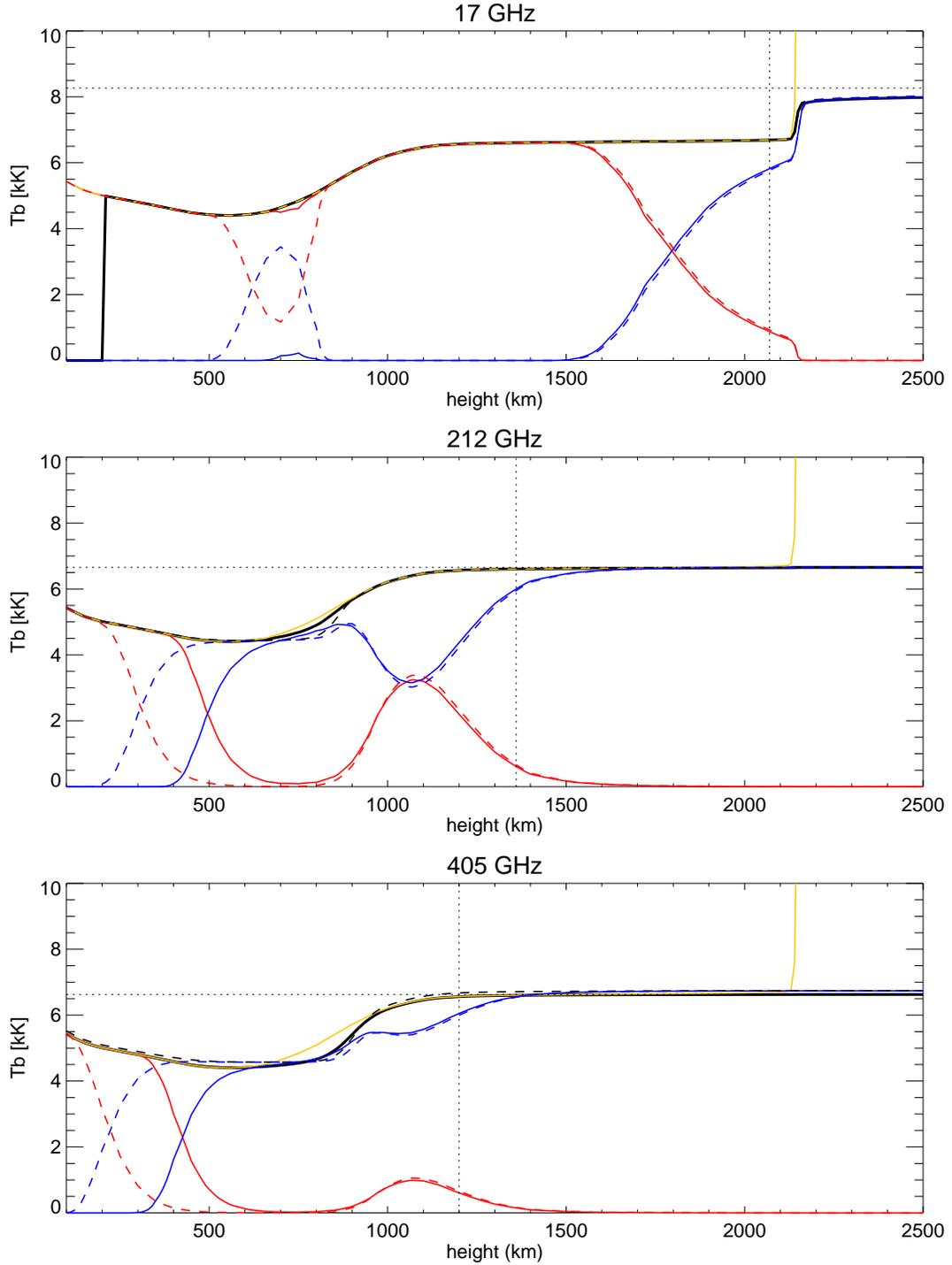}
\caption{Local absorption ($\epsilon_{abs}$, blue) and local emissivity 
($\epsilon_{emi}$, red) for 17, 212, and 405 GHz (top to bottom) 
using the C07 input atmospheric model. 
In yellow we 
show
the radial temperature and in black the brightness temperature.}\label{figure6.ps}

\end{center}
\end{figure}
\begin{figure}
\begin{center}
\includegraphics[width=1.0\textwidth]{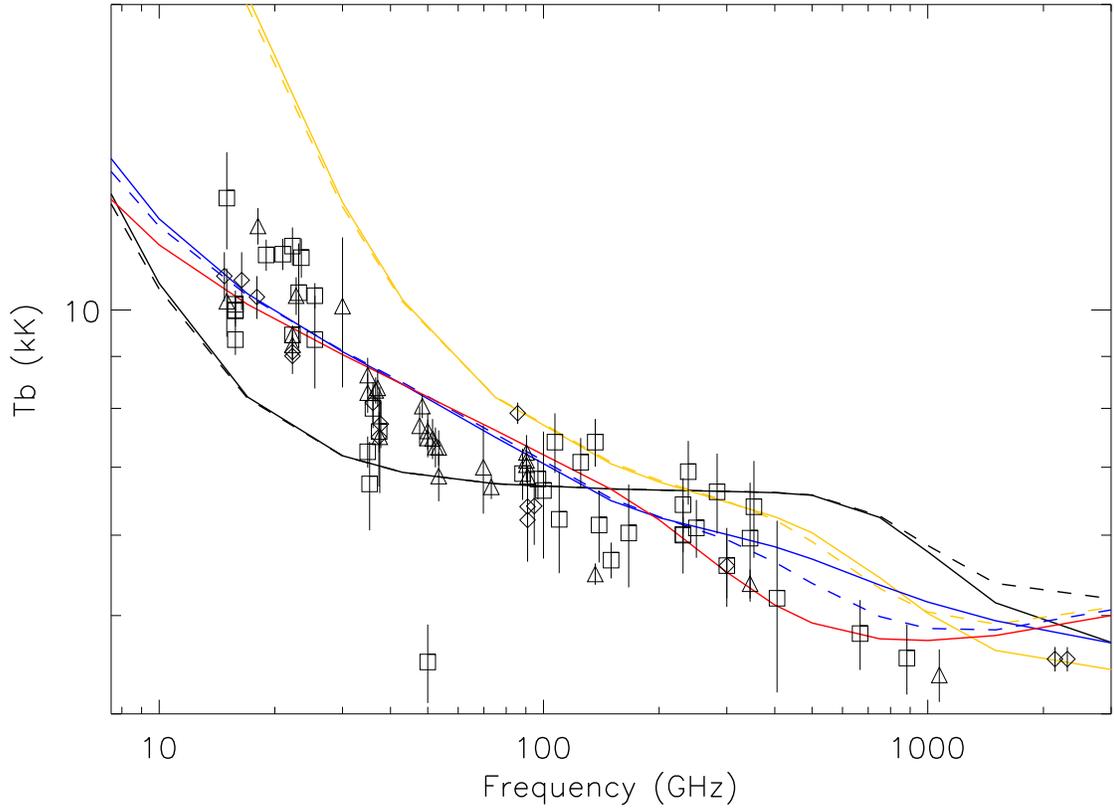}
\caption{Synthetic solar spectra and observations at different stages of 
solar activity cycle, collected by \cite{2004A&A...419..747L}. The square
points  
are intermediate, diamonds close-to-minimum, and triangles close-to-maximum of
solar activity cycle.  
The solar spectra are shown for bremsstrahlung opacity (dashed line) and
Celestun  
model (continuous line) for four input atmospheric models: C7 (black), SEL05
(blue), VALC (yellow) 
and SEL05 but using their electronic density (red).}\label{figure7.ps}
\end{center}
\end{figure}

At 212 and 405 GHz the behavior of the local emissivity changes noticeably, 
since 
at these frequencies two regions of emissivity are present: 
A lower one, closer to the
photosphere shows deep differences 
when comparing
between both models. 
The height of the region where the emission is generated is higher by
200 km when the Celestun model is used.
The uppermost 
region
is
located
between $800$ and $1500$ km above the photosphere. In this region there is no 
difference between the two models. 
As seen in Figure  \ref{figure6.ps}, the second layer of emission is below
the $\tau = 1$ height and is similar for both models. 
Even more, after $1000$ km, the brightness temperature remains unalterable.
This is due to the fact that 
the temperature of the atmospheric model 
is the upper limit for the brightness temperature, and remains constant between 
$1000$ and $2000$ km above the photosphere. 
the brightness temperature remains unalterable. 
These facts mask the local 
 emission processes which are  unnoticed by the observer.



We found that the dominant emission mechanism changes with altitude: 
$H^-$ is dominant below  
$500$ km whereas bremsstrahlung dominates at upper heights,
although the final brightness temperature
remains 
unchanged.
%


Our results show that the inclusion of 
 the $H^-$ and inverse bremsstrahlung mechanisms does not modify the
final brightness temperature.
But the $H^-$ opacity function 
%
has a significant impact in
the local emissivity and absorbency and it must be considered when studying 
micro structures in the low chromosphere. Therefore Celestun code is 
a valuable tool to perform this kind of studies.
%
%
The inverse bremsstrahlung 
has
no impact 
on
the
brightness temperature, neither on the local emissivity
nor on the
absorbency 
processes.

We note that there are more emission mechanisms acting at the low solar 
atmosphere, as instance, 
molecular emission has been reported  for CO \citep{1994Sci...263...64S} and
also 
 in the 515.60 to 516.20 nm spectral range  \citep[$C_2$ and $MgH$,
 ][]{2002A&A...382L..17F}. 
The 
emitting layer has been situated 
below the $1100$ km,
making evident
that these molecules are present in a
thin layer inside the temperature minimum region, between the
photosphere and the low chromosphere. 
However, 
observations in visible wavelengths, have shown that the molecular emission
at the limb brightening region is negligible, corresponding 
to  0.2\% of the continuum intensity at the center of the solar disk.
Nevertheless,
%
a detailed study of the
contribution of the molecular rotational spectrum at these wavelengths is
necessary.  


Although our simulations are related to quiescent
solar situations, the next natural steps will be to investigate
and compare with flaring cases. Therefore, such researches
could in principle be used to better understand the thermal
structure of the flaring solar atmosphere, that is the response of
the atmosphere to flare 
energy release. From the few existing
semi-empirical flare models like \cite{1980ApJ...242..336M} and
\cite{1990ApJ...360..715M} we should grossly expect 
decreasing radio 
fluxes
as a function of increasing wavelengths in the range
from centimeter to far-infrared wavelengths. 


We will pay special attention to
the density structure of the flaring loops (due to its importance
in the computed opacity functions). 
It is clear that new simulations, 
observations and their comparison are needed to better investigate the
structure and dynamics of the low layers of the solar atmosphere during
flares.

\acknowledgments
Part of this work was supported by UNAM-PAPPIT IN117309-3 and CONACyT 24879 
grants. Thanks to the National Center of Supercomputing in Mexico for allow us to use
his computer facilities and Dr. Emanuele Bertone for the comments. JPR thanks
CNPq agency (Proc.305655/2010-8). Thanks to Pr. Pierre Kaufmann, director of
CRAAM - Centro de Radioastronomia e Astrofísica Mackenzie, where part of this research was performed.

\bibliographystyle{apj}
\bibliography{libros}

\appendix

\end{document}